\documentclass[conference,letterpaper]{IEEEtran}

\IEEEoverridecommandlockouts
\usepackage{amsmath} 
\usepackage{amssymb} 
\usepackage{MnSymbol}
\usepackage{extarrows}
\usepackage{enumerate} 
\usepackage{bm} 
\usepackage{booktabs}
\usepackage{array}

\usepackage{graphicx} 
\usepackage{subfigure} 
\usepackage{multirow} 

\usepackage{cite}
\usepackage{amsmath,amsfonts}
\usepackage{algorithmic}
\usepackage{array}
\usepackage[caption=false,font=normalsize,labelfont=sf,textfont=sf]{subfig}
\usepackage{textcomp}
\usepackage{stfloats}
\usepackage{url}
\usepackage{verbatim}
\usepackage{xcolor}
\usepackage{graphicx}
\newtheorem{theorem}{Theorem}[section]
\newtheorem{definition}{Definition}[section]

\newtheorem{lemma}{Lemma}[section]

\newtheorem{proposition}{Proposition}[section]
\newtheorem{corollary}{Corollary}[section]
\newtheorem{remark}{Remark}[section]

\renewcommand{\P}{\mathbb{P}}

\newcommand{\pf}{\textbf{Proof: }}
\newcommand{\e}{\hfill$\blacksquare$}

\usepackage{epstopdf}

\allowdisplaybreaks[4] 

\date{}
\def\BibTeX{{\rm B\kern-.05em{\sc i\kern-.025em b}\kern-.08em
		T\kern-.1667em\lower.7ex\hbox{E}\kern-.125emX}}
\usepackage{balance}
\begin{document}

\title{Achievability Bounds on Unequal Error Protection Codes
}

%
%
%

\author{
	\IEEEauthorblockN{Liuquan Yao\IEEEauthorrefmark{1}\IEEEauthorrefmark{2},
		Shuai Yuan\IEEEauthorrefmark{1}\IEEEauthorrefmark{2},
		Yuan Li\IEEEauthorrefmark{3},
		Huazi Zhang\IEEEauthorrefmark{3},
		Jun Wang\IEEEauthorrefmark{3}
		Guiying Yan\IEEEauthorrefmark{1}\IEEEauthorrefmark{2}
		and Zhiming Ma\IEEEauthorrefmark{1}\IEEEauthorrefmark{2}}
	\IEEEauthorblockA{\IEEEauthorrefmark{1}%
		University of Chinese Academy and Sciences, Beijing, China}
	\IEEEauthorblockA{\IEEEauthorrefmark{2}%
		Academy of Mathematics and Systems Science, CAS, Beijing, China
	}
	\IEEEauthorblockA{\IEEEauthorrefmark{3}%
		Huawei Technologies Co. Ltd., China
	}
	Email: yaoliuquan20@mails.ucas.ac.cn, yuanshuai2020@amss.ac.cn,\\
	\{liyuan299, zhanghuazi, justin.wangjun\}@huawei.com, yangy@amss.ac.cn, 	mazm@amt.ac.cn
}

\maketitle

\begin{abstract}
	Unequal error protection (UEP) codes can
	facilitate the transmission of messages with different protection levels. In this paper, we study the achievability bounds on UEP by the generalization of Gilbert-Varshamov (GV) bound. For the first time, we show that under certain conditions,  UEP enhances the code rate comparing with time-sharing (TS) strategies asymptotically.
\end{abstract}
\section{Introduction}\label{section1}

\let\thefootnote\relax\footnotetext{\noindent This work is supported by National Key R\&D Program of China No. 2023YFA1009601 and 2023YFA1009602.}

\subsection{Unequal Error Protection}
In modern communication systems, simultaneous encoding of messages with disparate protection needs is essential. For
example, in certain 5G communication scenarios, it is necessary to transmit messages simultaneously for both Ultra-Reliable Low-Latency Communication (URLLC) and Enhanced Mobile Boradband (eMBB). A straightforward yet efficient strategy for achieving various error protection is time-sharing (TS). This approach involves independently encoding messages according to their required protection levels \cite{CC1}, \cite{CC2}, \cite{Multilevelcodes} \textit{etc}. Alternatively, the unequal error protection (UEP) technique encodes different messages as a whole, offering a distinct method for handling diverse reliability requirements.

The investigation on UEP codes can be categorized into two  areas. The first  pertains to the unequal protection of code bits  \cite{UBP1}, \cite{UBP2}.  In this paper, we are motivated by the need to transmit different types of messages and therefore focus on the other aspect: the unequal protection of information bits. The UEP for information bits was originally introduced in \cite{1978sep}, which centered on the linear unequal error protection (LUEP) codes.  \cite{1981}  provided a converse bound on the protection level for information bits of LUEP codes over the finite field $\mathbb{F}_q$ subsequently, which was proved to be achieved by Reed-Solomon codes. 
\cite{1981} also proposed  several practical strategies to enhance the protection capability, including code iterative and direct product.  These methods, together with direct sum and parity, were summarized in \cite{1984LUEP}. \cite{twotopic} derived achievability bounds and identified the minimum code length when the information bits are less than 14. However, calculating these bounds for longer codes is challenging. \cite{k-bounds} investigated the UEP for two different messages with minimum distance $2t+1$ and $3$, respectively. They
found the optimal non-binary LUEP solution in the sense of achieving the Hamming  bound based on Reed-Solomon code.

The general UEP, including nonlinear codes, has been conducted after 1990. Supposing the code length $n$  approaches infinity, several researches demonstrated that UEP codes do not achieve a higher error exponent compared to TS \cite{E}, \cite{E2}. Despite this, \cite{d} gave an asymptotic converse bound for the minimum  distance of UEP codes that exceeds the minimum distance achieved by TS. In the context of finite-length regime, \cite{Multilevelcodes}  introduced a strategy involving the partitioning of signal constellations into subsets. The critical data is encoded to specify which subset is selected, while the less important determines the exact signal in this subset. The underlying principle is that the distance between subsets is larger, hence the important data obtain more powerful protection. Several examples were also shown in \cite{Multilevelcodes} to enhance the minimum distance comparing with TS. However, these examples lack a corresponding theoretical result to  validate the improvement in error protection. Additionally, \cite{converse} presented a converse bound for two-level distance requirements.

Recent advancements in UEP codes  led to the development of several new techniques that enhance the protection capabilities. One such technique is based on the use of polar codes and the relations between information bits and coded bits \cite{polarUEP2018}, \cite{polar-uep}, \cite{polar3}, \cite{polar4} \textit{etc.} Furthermore, deep learning has also been explored to construct UEP codes \cite{DL}.

In this paper, we generalize the GV bound for binary UEP codes under multi-level protection requirement,   which is the first achievability bound that can be computed efficiently under arbitrary code lengths. In addition, for UEP codes with tow-level protection, we propose improved achievability bounds which better rate-distance trade-off.  Based on the proposed bounds, we provide several sufficient conditions to guarantee a non-vanishing rate improvement  over TS. Note that although our results are proved for the binary UEP codes, one can easily extend the binary case to general $q$-ary codes under similar arguments. 
%
%

Throughout this paper, we use bold font  $\bm{a}=(a_1,...,a_n)$   to represent vector, and abbreviate $(a_i,a_{i+1},\cdots,a_j)$ by $a_i^j$. We use $a\vee b = \max\{a, b\}$ to denote the maximum between $a$ and $b$.  The binary vector space is denoted by $\mathbb{F}_2^n$ and the Hamming distance is denoted by $d_H(\cdot,\cdot)$. 	We use $d_H(D_1, D_2)$ to  denote the Hamming distance between two sets $D_1$ and $D_2$, \textit{i.e.},
$
d_H(D_1, D_2):=\min_{c_1\in D_1, c_2\in D_2} d_H(c_1,c_2).
$
The Hamming ball in $\mathbb{F}_2^n$ centered at $x$ with radius $r$ is denoted by $B(x,r)$, and its volume is denoted by
$
V(n,r) := \sum\limits_{k=0}^r\binom{n}{k}.
$
The volume of the intersection of two Hamming balls with radius $r$ and center distance $d$ is denoted by
$
T(n,d,r) := \sum_{(s,t)\in O} \binom{d}{s}\binom{n-d}{t},
$
where $O = \{s,t\in \mathbb{Z}^+:s+t\leq r, t+d-s\leq r, s\leq d, t\leq n-d\}$. The  binary entropy function is denoted by $h(x)=-x\log_2x-(1-x)\log_2(1-x)$.

The rest of this paper is organized as follows.
The definitions of UEP codes and GV bound are introduced in Section \ref{section2}, while the main theorem  for UEP codes is provided in Section \ref{section3}. Section \ref{section4} establishes the comparison between TS and   UEP codes, and  simulation results are presented in Section \ref{section5}. We draw the conclusion in Section \ref{section6}.

%
%
%

\section{Preliminaries}\label{section2}
We first define the (binary) UEP codes with different distance requirement.
\begin{definition}\label{de of UEP}
	Given $m$ message sets $\mathcal{A}_1, \mathcal{A}_2,\cdots,\mathcal{A}_m$ with $|\mathcal{A}_i|=A_i, i=1,...,m$. An $(A_1^m, d_1^m)$ unequal error protection code on $\mathbb{F}_2^n$ is a map
	\begin{equation}\label{map}
	\begin{aligned}
	C:\; &\mathcal{A}_1\times \mathcal{A}_2\times\cdots\times\mathcal{A}_m\rightarrow \mathbb{F}_2^n,\\
	&(a_1,a_2,\cdots,a_m)\mapsto c\in \mathbb{F}_2^n,
	\end{aligned}
	\end{equation}
	such that for all $i=1,2,\cdots, m$,
	\begin{equation}
	d_i=\min\{d_H(c,c'): c, c'\in C, (C^{-1}(c))_i\neq (C^{-1}(c'))_i   \},
	\end{equation}
	where we still use alphabet $C$ to denote the code book of map $C$. We call $A_i$ the size of code for message set $\mathcal{A}_i$ and $d_i$ the minimum distance for the message set $\mathcal{A}_i$.
	
\end{definition}
\begin{remark}
	Let $m=n$ and $\mathcal{A}_i=\{0,1\},\forall i=1,2,\cdots,n$, then Definition \ref{de of UEP} is coincided with the UEP codes defined in   \cite{1981}, \cite{twotopic} and \cite{Multilevelcodes} \textit{etc.} Further, the vector $\bm{d}=(d_1,d_2,\cdots,d_n)$ is called the separate vector or information distance profiles, which implies the protection levels of information bits \cite{1978sep}, \cite{1981}, \cite{disp}.	
\end{remark}

The classic GV bound provides an achievable size of message given the code length and minimum distance.
\begin{lemma}[GV bound]\label{GV bound}
	There exists a binary code $C$ with length $n$, minimum distance $d$ and size
	$
	|C| \geq \frac{2^{n}}{V(n,d-1)}.
	$
\end{lemma}
By considering the intersection of Hamming balls, \cite{inter} derived an improved GV bound as follows.
\begin{lemma}[Improved GV bound \cite{inter}]\label{improved GV bound}
	There exists a binary code $C$ with length $n$, minimum distance $d$ and size
	\begin{equation}\label{G}
	|C| \geq \frac{2^{n}-T(n,d,d-1)}{V(n,d-1)-T(n,d,d-1)}:=G(n,d).
	\end{equation}
\end{lemma}
The enhancement stems from considering the intersection of the balls.
Specifically, Lemma \ref{GV bound} is from the \textbf{union bound}
\begin{equation}\label{union}
|\bigcup_{i=1}^M B(c_i,d-1)|\le MV(n,d-1),
\end{equation}
while Lemma \ref{improved GV bound} suggests to construct codewords $c_1,c_2,\cdots,c_M$ with code distance exactly $d$, \textit{i.e.},  for any $c_i, i>1$, there exists $c_j, j<i$ \textit{s.t}
\begin{equation}\label{d=d}
d_H(c_i,c_j)=d.
\end{equation}
Considering the intersection of $B(c_i,d-1)$ and $B(c_j,d-1)$, we have
\begin{equation}\label{M-T}
|\bigcup_{i=1}^M B(c_i,d-1)|\le MV(n,d-1)-(M-1)T(n,d,d-1),
\end{equation}
which is  tighter than \eqref{union}.
\begin{figure}[!t]
	\centering
	\includegraphics[width=0.2\textwidth]{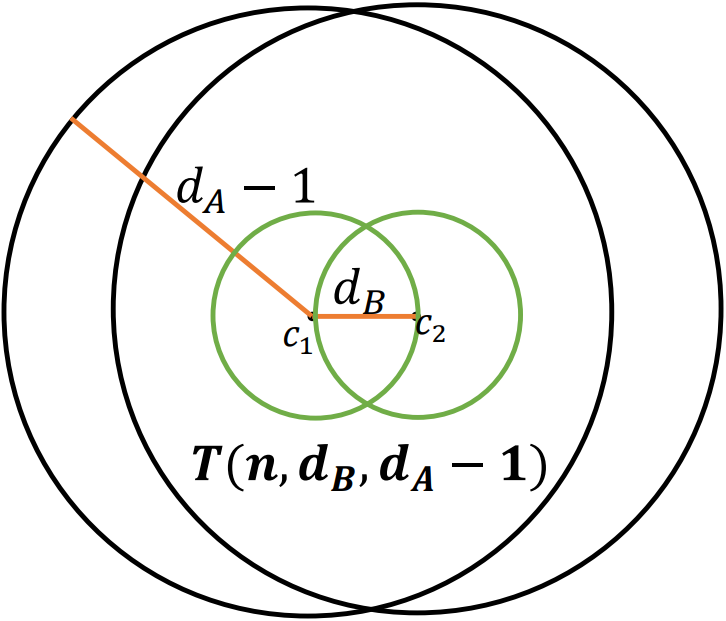}
	\caption{Intersection Bound}
	\label{fig of volume}
\end{figure}

In the next section,  we  generalize the GV bounds to UEP codes with multi-level protection capabilities.
\section{Main Results}\label{section3}
	In section \ref{3.A} we generalize the GV bound to get an achievability bound for multi-level protection UEP codes (Theorem \ref{GV-main}).  An improved achievability bound for two-level protection UEP codes (Theorem \ref{Best theorem}) is derived in Section \ref{3.B}, in which Proposition \ref{tight-union-bound} introduces the intersection bounds for estimating volume of union balls, and Proposition \ref{prop of CS}  provides a more relaxed sufficient condition for Proposition \ref{tight-union-bound}. In section \ref{3.C} we consider another way   to improve the achievability bound for two-level protection UEP codes, called the enlargement bound (Theorem \ref{at least MS}).
	\subsection{GV Bound for UEP Codes}\label{3.A}
We utilize the union bound \eqref{union} and a greedy strategy to select codewords to establish an achievability bound for UEP codes  in the following theorem.

\begin{theorem}\label{GV-main}
	Given $m$ message sets $\mathcal{A}_1, \mathcal{A}_2,\cdots,\mathcal{A}_m$ with $|\mathcal{A}_i|=A_i$, and $ d_1\ge d_2\ge \cdots\ge d_m$.
	Suppose a subset $ W\subset \mathbb{F}_2^n$ satisfies $|W|>S(n,m,A_1^m, d_1^m )$, where
	\begin{equation}
	S(n,m,A_1^m, d_1^m ):=\sum_{i=1}^m \left( \prod_{j>i}A_j\right) (A_i-1)V(n,d_i-1),
	\end{equation}
	then there exists an   $(A_1^m, d_1^m)$ UEP code $C$ of message $\mathcal{A}_1\times \mathcal{A}_2\times\cdots\times\mathcal{A}_m$  such that
	$
	C\subset W.
	$
\end{theorem}
\pf We prove Theorem \ref{GV-main} via induction on $m$. If $m=1$, then the statement is true due to the classic GV bound. For the induction step $m-1 \rightarrow m$, take $a\in \mathcal{A}_1$ and split the message set into
$
\mathcal{M}_1:=\{a\}\times\cdots\times\mathcal{A}_{m-1}\times \mathcal{A}_m,\;\; \mathcal{M}_2:=(\mathcal{A}_1\setminus\{a\})\times\cdots\times\mathcal{A}_{m-1}\times \mathcal{A}_m.
$
The codewords are greedily selected as follows.
\begin{enumerate}
	\item We first choose codewords for $\mathcal{M}_2=\{m_1,m_2, \cdots,m_{M_2}\}$. Take arbitrary $c_1\in W$ as the codeword for $m_1$.
	\item Suppose we have encoded $\{m_1,...,m_{i}\}$ into $\{c_1,...,c_{i}\}$ for some $ 1\leq i<M_2 = (A_1-1)( \prod_{k=2}^{m}A_k)$. Since
	\begin{align*}
	 	|W|-(i+1)V(n,d_{1}-1)&>|W|-(M_2+1)V(n,d_{1}-1)\\
	 	&>|W|-S(n,m,A_1^m,d_1^m)>0,
	\end{align*}
	we can choose a codeword $c_{i+1}\in W$ for $m_{i+1}$ such that $d_H(c_i,c_j)\ge d_1,\;\forall j<i+1$. Continue this procedure until the whole $\mathcal{M}_2$ are encoded into $\bar{C} = \{c_1,c_2,...,c_{M_2}\}$. Note that $d_1\geq d_i$ for all $i>1$, which implies $\bar{C}$ is an $((A_1-1, A_2^{m}),d_1^m)$ UEP code for $\mathcal{M}_2$.
	
	\item Next we select codewords for messages in $\mathcal{M}_1$. Let
	$
	C_1:=W\setminus \bigcup_{i=1}^{M_2}B(c_i, d_1-1).
	$
	By union bound \eqref{union}, we have
	\begin{equation}\label{reason-for-loose}
	\begin{aligned}
	&|C_1|\ge |W|-|\bigcup_{i=1}^{M_2}B(c_i, d_1-1)|\\
	&\ge |W|- ( \prod_{i=1}^{m-1}A_i) (A_m-1)V(n,d_m-1)\\
	&>S(n,m,A_1^{m},d_1^{m})-( \prod_{i=1}^{m-1}A_i) (A_m-1)V(n,d_m-1)\\
	&= S(n,m-1,A_2^{m},d_2^{m}),
	\end{aligned}
	\end{equation}
	and thus by induction assumption, we can find an $(A_2^{m},d_2^{m})$ UEP code $\tilde{C}\subset C_1$ for $\mathcal{M}_1$, since this is code for $m-1$ message sets.
	
	\item Let $C = \bar{C} \cup \tilde{C}$. According to the construction of $C_1$ we know that $C$ is an $(A_1^m,d_1^m)$ UEP code, and this completes the proof.
\end{enumerate}
\e
\begin{remark}
	 If $d_1=d_2=\cdots,d_m$, then Theorem \ref{GV-main} coincides with the standard GV bound \ref{GV bound}.
\end{remark}
The applicability of Theorem \ref{GV-main} extends beyond the binary field  to $\mathbb{F}_q$. The method outlined in Theorem \ref{GV-main}  serves as a foundational concept that can be further enhanced and refined in the next section.

\subsection{Improved Bound for Two-level Protection}\label{3.B}
Let $\mathcal{A}=\{m^\mathcal{A}_1,m^\mathcal{A}_2,\cdots, m^\mathcal{A}_A\}$ and $\mathcal{B}=\{m^\mathcal{B}_1,m^\mathcal{B}_2,\cdots,m^\mathcal{B}_B\}$ be two message sets with $|\mathcal{A}|=A$ and $|\mathcal{B}|=B$.  Consider a two-level protection UEP code of length $n$ with minimum distance $d_A$ and $d_B$ for $\mathcal{A}$ and $\mathcal{B}$, respectively, and the rates of message $\mathcal{A}$ and $\mathcal{B}$ are defined as $R_A:=\frac{\log_2 A}{n}$ and $R_B:=\frac{\log_2 B}{n}$. For convenience, we denote such code by $(n,A,B,d_A,d_B)$-UEP code. We assume that $d_A > d_B$, \textit{i.e.}, the message set $\mathcal{A}$ requires stronger error protection ability than $\mathcal{B}$. By Theorem \ref{GV-main}, we   obtain the following achievable size of $\mathcal{A}$ with given $n,B,d_A,d_B$.
\begin{corollary}\label{GV-2-1}
	There exists an $(n,A,B,d_A,d_B)$-UEP code for the message set $\mathcal{A}\times \mathcal{B}$ with
	\begin{equation}\label{union size}
	A\geq \frac{2^n-(B-1)V(n,d_B-1)}{BV(n,d_A-1)}.
	\end{equation}
\end{corollary}
\begin{remark}\label{LUEP}
Following the line of proof demonstrating that the GV bound can be attained by linear binary codes \cite{LGV}, we can similarly establish that the size specified in Corollary \ref{GV-2-1} can be achieved by LUEP codes, see details in Appendix \ref{pf of LUEP}.
\end{remark}

When $d_B\ll d_A$, the achievability bound provided by Corollary \ref{GV-2-1} is not  tight.
This is primarily due to the inefficiency of the union bound. In fact, if we gather the codewords $c_i$ with code distance $d_B$, the volume of the intersection of the balls $B(c_i,d_A-1)$ will be large, but the union bound does not take this into account, which results in a less efficient utilization of the space.
In view of the above analysis, we  provide an improved GV bound for the two-level protection UEP codes by deriving a tighter estimation, which is a generalization version of \eqref{M-T}  by changing $d-1$ to arbitrary $r\ge 0$..
\begin{proposition}\label{tight-union-bound}
	Let $\mathcal{C}=\{c_1,c_2,...,c_N\}$ be a code such that for any $i=2,3,...,N$, there exists $j < i$ \textit{s.t.} $d_H(c_i,c_j) = d$. Then
	\begin{equation}\label{I}
	|\bigcup_{k=1}^N B(c_k,r)| \leq NV(n,r) - (N-1)T(n,d,r):=I(n,N,d,r).
	\end{equation}
\end{proposition}
\pf Denote $B_k = B(c_k,r)$ and $\tilde{B}_k = B_k\setminus (\cup_{i<k}B_i),k=1,2,...,N$, then
$
|\bigcup_{k=1}^N B(c_k,r)| = |\bigcup_{k=1}^N \tilde{B}_k| = \sum\limits_{k=1}^N |\tilde{B}_k|.
$
For each $k\geq2$, let $j(k)$ be the index such that $d_H(c_{j(k)},c_k)=d$. It follows that
\begin{equation}
\begin{aligned}
\sum\limits_{k=1}^N |\tilde{B}_k|&\leq |B_1| + \sum\limits_{k=2}^N |B_k\setminus B_{j(k)}| \\
&= V(n,r) + (N-1)(V(n,r)-T(n,d,r)).
\end{aligned}
\end{equation}
\e

	We call \eqref{I}  the \textbf{intersection bound} and Fig.\ref{fig of volume} exhibits an example with $d=d_B, r=d_A-1$. To ensure the existence of the code satisfying the conditions in Proposition \ref{tight-union-bound}, in the following we introduce the concept of connected set.
\begin{definition}[Connected Set]
	Let $x,y\in\mathbb{F}_2^n$, a path of length $t$ between $x$ and $y$ is $t+1$ ordered points $\{x_0,x_1,...,x_t\}\subset\mathbb{F}_2^n$ such that $x_0=x,x_t=y$ and $d_H(x_i,x_{i+1})=1,i=0,1,...,t-1$. A subset $D\subset \mathbb{F}_2^n$ is called a connected set if for any $x,y\in D$, there exists a path $\{x_0,x_1,...,x_t\},t\ge 0$ between $x$ and $y$ where all $x_i\in D,i=0,1...,t$.
\end{definition}

The next proposition shows that we can construct codewords   on a connected set to satisfy the condition  in Proposition \ref{tight-union-bound}.
\begin{proposition}\label{prop of CS}
	Let $D\subset \mathbb{F}_2^n$ be a connected set with $|D|>(B-1)V(n,d_B-1)$, then there exists a code $C=\{c_1,c_2,\cdots,c_B\}\subset D$ with size $B$ and minimum distance $d_B$ such that
	\begin{equation}\label{compact codes}
	\forall i\leq B, \exists j<i,\ s.t.\ d_H(c_i,c_j)=d_B.
	\end{equation}
\end{proposition}
\pf 
Take $c_1\in D$ arbitrarily. Suppose we have selected $j$ codewords $\{c_1,c_2,...,c_j\}$ for some $1\leq j <B$. Let $W = \cup_{i=1}^j B(c_i,d_B-1)$, then $D\setminus W$ is non-empty since $|D|> jV(n,d_B-1)-(j-1)T(n,d_B,d_B-1)\ge |W|$, according to Proposition \ref{tight-union-bound}. For $x,y\in D$, let $\ell(x,y)$ be the length of the shortest path between $x$ and $y$. Define
\begin{equation}
r = \min\limits_{x\in D\setminus W , y\in W} \ell(x,y),
\end{equation}
and suppose $r$ is achieved by $x'\in D\setminus W$ and $y'\in W$. We claim that $r=1$. In fact, let $\{x'=x_0,x_1,...,x_{r-1},x_r=y'\}\subset D$ be the shortest path between $x'$ and $y'$. If $r>1$, then there would be a contradiction that neither $x_1$ can belong to $D\setminus W$ or we would find a shorter path $\{x_1,...,x_{r-1},y'\}$ connecting $D\setminus W$ and $W$, nor it can belong to $W$ otherwise it would be possible to find another shorter path $\{x',x_1\}$. Now it follows that
$
d_H(x',y') \leq \ell(x',y') = r= 1,
$
since Hamming distance between two points is equal to the length of the shortest path that connects them in $\mathbb{F}_2^n$. Take $c_i$ such that $d_H(c_i,y')\leq d_B-1$, then
\begin{equation}
d_H(c_i,x') \leq d_H(c_i,y') + d_H(y',x') \leq d_B.
\end{equation}
On the other hand we have $d_H(c_i,x') \geq d_B$ because $x'\in D\setminus V$. This implies $d_H(c_i,x')=d_B$, and we can take $c_{j+1} = x'$ as the $j+1$-th selected codeword. Clearly the codewords selected by the above procedure satisfy the desired property.\e

	Given $n,B,d_A,d_B$,  define $\mathcal{D}(n,B,d_A,d_B)$ to be the collection of all subsets families $\{D_1,D_2,...,D_M\}$ where $D_i\subset \mathbb{F}_2^n, \forall i=1,2,\cdots, M$ such that
\begin{enumerate}
	\item $D_i$ is a connected set, $\forall i$.
	\item $|D_i|>(B-1)V(n,d_B-1),\forall i$.
	\item $d_H(D_i,D_j)\ge d_A, \forall j\neq i$.
\end{enumerate}
Suppose $\{D_1,D_2,...,D_M\}\in\mathcal{D}(n,B,d_A,d_B)$. For each $D_i$, we can associate $m_i^\mathcal{A}\times \mathcal{B}$ with $B$ codewords $\{c_{ij}\}_{j=1}^B$ in $D_i$ with minimum distance $d_B$ satisfying \eqref{compact codes} by code map $C_i$. And 3) allows us to guarantee the distance between codewords in $C_i(m_i^\mathcal{A}\times \mathcal{B})\subset D_i$ and  $C_j(m_j^\mathcal{A}\times \mathcal{B})\subset D_j$ with $i\neq j$ is larger than $d_A$. Let $\bar{C} = \{c_{ij}: 1\leq i\leq M, 1\leq j \leq B\}$ and $B_U = \bigcup_{i,j} B(c_{ij},d_A-1)$. Taking $W = \mathbb{F}_2^n\setminus B_U$ and $m=2$. By Theorem \ref{GV-main}we know that there exists an   $(n,A',B,d_A,d_B)$-UEP code $\tilde{C}\subset W$ if $|W|>S(n,2,(A',B),(d_A,d_B))=B(A'-1)V(n,d_A-1)+(B-1)V(n,d_B-1)$. Or equivalently, we can find such code with
\begin{equation}
\begin{aligned}\label{A'}
&A' \geq \frac{|W| - (B-1)V(n,d_B-1)}{BV(n,d_A-1)}\vee 0.
\end{aligned}
\end{equation}
Let $C = \bar{C}\cup\tilde{C}$, then clearly $C$ is an $(n,A,B,d_A,d_B)$-UEP code with
\begin{equation}\label{tight-GV-bound}
\begin{aligned}
&A = A' + M\\
&\overset{(a)}{\geq}M+
\frac{2^n - (B-1)V(n,d_B-1)-MI}{BV(n,d_A-1)}\vee 0,\\
\end{aligned}
\end{equation}
where $(a)$ follows from the estimation of $|W|$ using Proposition \ref{tight-union-bound} and \eqref{I}, $I=I(n,B,d_B,d_A-1)$. \eqref{tight-GV-bound} provides a tighter achievability bound compared with Corollary \ref{GV-2-1}, which is summarized   in the next theorem.
\begin{theorem}\label{Best theorem}
	Let
	\begin{equation}
	M^* = \sup\{M:\exists \{D_1,...,D_M\}\in\mathcal{D}(n,B,d_A,d_B)\},
	\end{equation}
	then there exists an $(n,A,B,d_A,d_B)$-UEP code such that
	\begin{equation}\label{M*}
	\begin{aligned}
	&A\geq  M^*+
	\frac{2^n - (B-1)V(n,d_B-1)-M^*I}{BV(n,d_A-1)}\vee 0,
	\end{aligned}
	\end{equation}
	where 
	$
	I=I(n,B,d_B,d_A-1).
	$
\end{theorem}

The analysis of $M^*$ is challenging. However, we can take a specific choice of $\{D_1,...,D_M\}\in\mathcal{D}(n,B,d_A,d_B)$ to obtain a lower bound on $M^*$.

One natural choice of $D_i$s is cubes, which implies that the achievable size in Theorem \ref{Best theorem} is always better than TS (see details in Section \ref{3.2}).
Another  choice of $D_i$s is the Hamming balls with radius  
$
r_V=\min\{r: V(n,r)> (B-1)V(n,d_B-1)\}.
$
In order to ensure the distance of $D_i$s to be not less than $d_A$, we pack the larger disjoint Hamming balls $B_i$ with radius $r_S = r_V + \lceil\frac{d_A}{2}\rceil$ in $\mathbb{F}_2^n$, and select the concentric balls of $B_i$ with radius $r_V$ as $D_i$.


Let $M_S$ be the maximum number of disjoint Hamming balls with radius $r_S$, using the classic GV bound and Hamming bound, we have
\begin{equation}\label{bound of MS}
\frac{2^n}{V(n,2r_S)} \leq M_S \leq \frac{2^n}{V(n,r_S)}.
\end{equation}
Consequently, $M^*\geq M_S\geq 2^n / V(n,2r_S)$, which provides a computable lower bound on $M^*$.

\subsection{Enlargement Union Bound}\label{3.C}
\begin{figure}[!t]
	\centering
	\includegraphics[width=0.35\textwidth]{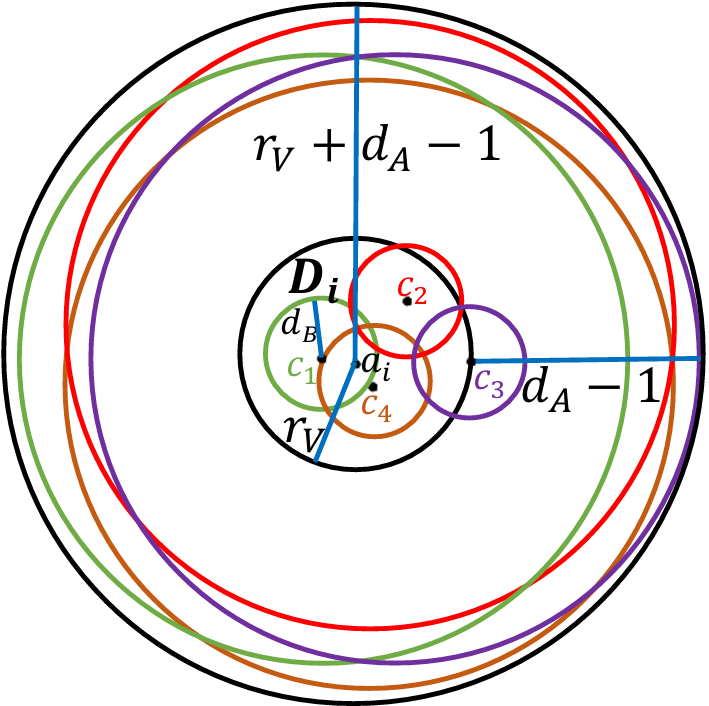}
	\caption{Enlargement Bound}
	\label{enlargen fig}
\end{figure}

If we choose $D_i$s as balls, another volume estimation of the union ball can be built as follows,
\begin{equation}\label{enlarge}
|\bigcup_{k=1}^B B(c_k,d_A-1)|\le V(n,r_V+d_A-1),
\end{equation}
where $c_1,c_2,\cdots,c_B$ belong to a ball with radius $r_V$, as shown in Fig.\ref{enlargen fig}. We call \eqref{enlarge} the \textbf{enlargement bound}.  The intersection bound \eqref{I} only considers the  intersection between two balls, which is loose when the balls are  dense. The   enlargement bound  considers the intersections of all the balls, which is beneficial when $d_B\ll d_A$.

Since the total space $\mathbb{F}_2^n$ is connected,  we can find disjoint balls $D_1,D_2,\cdots,D_{M_S}$ with centers $a_1,a_2,\cdots,a_{M_S}$ with radius $r_V(<r_S)$ satisfying
\begin{equation}
\forall i\le M_S,\;\; \exists j<i, \;\;\textit{s.t.}\;\; d_H(a_i,a_j)=2r_S+1,
\end{equation}
by the following proposition.
\begin{proposition}
	If $\mathbb{F}_2^n$ can obtain $M$ disjoint balls with radius $r$, then there exists disjoint balls $D_1,D_2,\cdots,D_{M}$ with centers $a_1,a_2,\cdots,a_{M}$ and radius $r$ satisfying
	\begin{equation}
	\forall i\le M,\;\; \exists j<i, \;\;\textit{s.t.}\;\; d_H(a_i,a_j)=2r+1.
	\end{equation}
\end{proposition}
\pf We denote the set of $M$ disjoint balls in $\mathbb{F}_2^n$ with radius $r$ as $E_0$, and do the following steps.
\begin{enumerate}
	\item  Choose one ball in $E_0$ arbitrarily and denote it as $D_1$ with center $c_1$. Denote $\{D_1\}=F$, $E=E_0\setminus F$. 
	
	\item If the distances between all centers of balls in $F$  and all centers of balls in $E$ are  larger than $2r+1$, then by triangle inequality,
	\begin{equation}
	d_H(F, E)>1.
	\end{equation}
	Find $x\in F, y\in E$ such that $d_H(x,y)=d_H(F, E)$ and find the shortest path between $x$ and $y$ as $\{x,x_1,x_2,\cdots,x_t,y\}$, obviously, $t=d_H(F, E)-1$, and $d_H(x_i,x)=i$. Then we define
	\begin{equation}
	G=F+(x_t-x):=\{z\in \mathbb{F}_2^n: z-(x_t-x)\in F\},
	\end{equation}
	and change $D_i$s to $D_i+(x_t-x)$s and still denote them as $D_i$s.
	Clearly  $d_H(G,y')\ge 1, \forall y'\in E$ otherwise there exists contradiction for the definition of $(x,y)$, and $d_H(G,y)=1$. Since $y$ belongs to a ball $\bar{D}\in E$ with center $\bar{c}$, and $x$ belongs to a ball $\hat{D}\in G$ with center $\hat{c}$, by triangle inequality we conclude that
	\begin{equation}
	d_H(\hat{c},\bar{c})=2r+1.
	\end{equation}

	\item If there exists a ball in $E$ and a ball in $F$ such that their centers have distance $2r+1$, then we take $G=F$ directly.

    \item
	Find all balls in $E$ such that their centers have distance $2r+1$ with one of the center of the balls in $G$,  and denote them as $E^i$. Give the numbers for balls in $E^i$ as $D_{|G|+1},\cdots,D_{|G|+|E^i|}$, and do assignments $E_0\leftarrow E, F\leftarrow G\cup E^i, E\leftarrow E_0\setminus E^i$.
	
	\item Do step 2-4 until $E_0=E^i$.
\end{enumerate}
Clearly, the balls in final $G$ satisfy the requirements.
\e

	Therefore, we deduce 
\begin{equation}
\begin{aligned}
&|\bigcup_{1\le i\le M_S,1\le j\le B}B(c_{ij},d_A-1)|
\le |\bigcup_{i=1}^{M_S} B(a_{i},r_V+d_A-1)|\\
&\le I(n,M_S,2r_S+1,r_V+d_A-1),
\end{aligned}
\end{equation}
where $\{c_{ij},1\le j\le B\}\subset D_i, i=1,2,\cdots,M_S$.
Clearly,   another achievability bound of UEP  is obtained as follows.
\begin{corollary}\label{GV-ball}
	There exists an $(n,A,B,d_A,d_B)$-UEP code for the message set $\mathcal{A}\times \mathcal{B}$ with
	\begin{equation}\label{enlarge size}
	A\geq M_S+\frac{2^n-I-(B-1)V(n,d_B-1)}{BV(n,d_A-1)}\vee 0,
	\end{equation}
	where $I=I(n,M_S,2r_S+1,r_V+d_A-1)$.
\end{corollary}

The following theorem describes the gains of code size $A$ from union bound to enlargement bound.
\begin{theorem}\label{at least MS}
		Given $n,B,d_A,d_B$ satisfying
	\begin{equation}
	d_B<d_A,\;\; BV(n,d_A)\le 2^n.
	\end{equation}
	Denote the achievable sizes \eqref{union size} and \eqref{enlarge size} as $A^{union}$ and $A^{enlarge}$ respectively. If 
	\begin{equation}\label{d_B small}
		h\left( \frac{d_A}{n}+h^{-1}\left( R_B+h(\frac{d_B}{n})\right) \right)<R_B+h(\frac{d_A}{n}),
	\end{equation}
	where   $R_B=\frac{\log_2 B}{n}$,
	then
	there exists $\gamma=\gamma(B,d_A,d_B)\in(0,1)$ \textit{s.t.}
	\begin{equation}\label{M_S}
	A^{enlarge}-A^{union}\ge M_S(1-O(\gamma^n)).
	\end{equation}
	If we further suppose that
	\begin{equation}\label{condition of rate improve}
	h\left( \frac{d_A}{n}+2h^{-1}\left( R_B+h(\frac{d_B}{n})\right) \right)<R_B+h(\frac{d_A}{n}),
	\end{equation}
	then there exists $\Gamma=\Gamma(B,d_A,d_B)>1$ \textit{s.t.}
	\begin{equation}\label{rate improve}
	\frac{A^{enlarge}}{A^{union}}=O(\Gamma^n).
	\end{equation}
\end{theorem}
\pf 
Since we assume  $BV(n,d_A)<2^n$, conditions \eqref{d_B small} and \eqref{condition of rate improve} ensure all volume of balls in the remaining proof can be approximated by the following well-known estimation (\textit{e.g.}, see \cite{yehezkeally2022bounds}):
\begin{equation}
\frac{1}{n+1}2^{n h(r/n)}\le V(n,r)\le 2^{n h(r/n)}, \forall r\leq \frac{n}{2}.
\end{equation}

Now  we prove \eqref{M_S} and \eqref{rate improve}.   The symbol $\approx$ is employed denote that the values on both sides of ``$\approx$'' are sufficiently closed when the code length is large enough.

 First we estimate $r_V$ by
\begin{equation}
h(\frac{r_V}{n})\approx R_B+h(\frac{d_B}{n})
\end{equation}
when $n$ is large and thus $\frac{r_V}{n}\approx h^{-1}(R_B+h(\frac{d_B}{n})):=\eta$, then
\begin{align}
&A^{enlarge}-A^{union}=M_S-\frac{I(n,M_S,2_S+1,r_V+d_A-1)}{BV(n,d_A-1)}\\
&\ge M_S\left( 1-\frac{V(n,r_V+d_A-1)}{BV(n,d_A-1)}\right) \approx M_S\left( 1-\frac{V(n,r_V+d_A)}{BV(n,d_A)}\right) \\
&\ge M_S\left( 1-\frac{(n+1)2^{nh(\frac{r_V+d_A}{n})}}{2^{n(R_B+h(\frac{d_A}{n}))}}\right)\\
&\approx M_S\left( 1-(n+1)2^{-n(R_B+h(\frac{d_A}{n})-h(\frac{d_A}{n}+\eta)}\right).
\end{align}

According to the above proof, we have
\begin{align}
&\frac{A^{enlarge}}{A^{union}}\approx \frac{M_S}{2^n/(BV(n,d_A))}\\
&\ge \frac{2^{n(R_B+h(\frac{d_A}{n}))}}{2^nh(\frac{2r_V+d_A}{n})}\\
&\approx \frac{2^{n(R_B+h(\frac{d_A}{n}))}}{2^nh(2h^{-1}\left( R_B+h(\frac{d_B}{n})\right)+\frac{d_A}{n})}\\
&=\left( 2^{[R_B+h(\frac{d_A}{n})-h(2h^{-1}\left( R_B+h(\frac{d_B}{n})\right)+\frac{d_A}{n})]}\right)^n.
\end{align}
The proof completes.
\e

\begin{remark}
\eqref{d_B small} holds when 
\begin{equation}
g\left( h^{-1}(R_B),\frac{d_A}{n}\right)\ge h(\frac{d_B}{n}).
\end{equation}
where $g(x,y):=h(x)+h(y)-h(x+y)$. Since $g(x,y)>0$ (see Appendix \ref{g>0}), the condition \eqref{d_B small} can be satisfied if $\frac{d_B}{n}$ is small enough, and the ascension of code size $A$ generated by enlargement bound  is exponential as n increases, comparing with the one by the union bound. 	Under \eqref{condition of rate improve}, the enlargement bound is able to provide non-vanishing rate improvement asymptotically. Several examples satisfying \eqref{condition of rate improve} are shown in the following:
\begin{enumerate}
	\item $\frac{d_B}{n}=0.001, B=2^{0.7n}, \frac{d_A}{n}\in[0.051,0.053]$.
	
	\item $\frac{d_B}{n}=0.01, B=2^{0.5n}, \frac{d_A}{n}\in[0.094,0.110]$.
\end{enumerate}
\end{remark}

\begin{figure}[!t]
	\centering
	{		\subfigure
		{\includegraphics[width=0.24\textwidth]{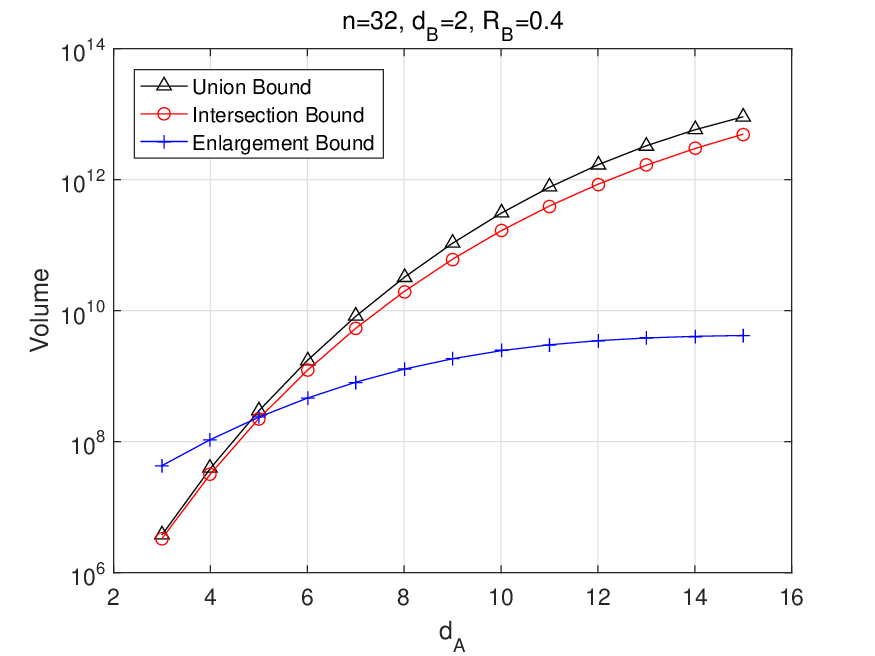}}
		\subfigure
		{\includegraphics[width=0.24\textwidth]{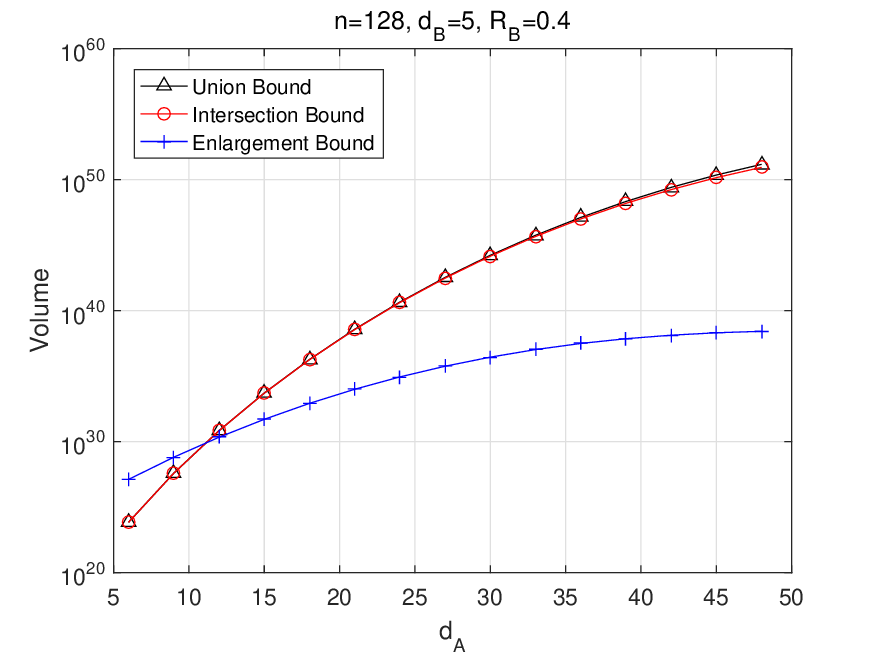}}
	}
	\caption{Comparison of  the Volume Bounds of the Union of Balls}
	\label{Comparison of Bounds}
\end{figure}

	We show a comparison between union bound \eqref{union}, intersection bound \eqref{I} and enlargement bound \eqref{enlarge}  in Fig. \ref{Comparison of Bounds}. Apparently, when $d_B\ll d_A$, the enlargement bound   is significantly tighter compared to  other  bounds.

\section{Comparison between UEP and Time-sharing}\label{section4}
\subsection{Time-sharing codes}
The TS codes are the simplest UEP codes, and we consider two message sets $\mathcal{A}$ and $\mathcal{B}$ with size $A^{TS}$ and $B^{TS}$, respectively. Define the $(n_A, n_B, A^{TS},B^{TS},d_A,d_B)$ TS code as the combination of two codes
\begin{equation}
C^A_{TS}:\mathcal{A}\to F_2^{n_A},\;\; C^B_{TS}:\mathcal{B}\to F_2^{n_B},
\end{equation}
where the minimum distance of $C_{TS}^A$ and $C_{TS}^B$ are $d_A$ and $d_B$, respectively. In other word, the TS codes encode the message sets $\mathcal{A}$ and $\mathcal{B}$ independently using $C_{TS}^A$ and $C_{TS}^B$, respectively.
The following bounds for TS codes can be easily obtained by the classic GV bound. 
\begin{lemma}[GV bound for TS codes]
	There exists an $(n_A,n_B,A^{TS},B^{TS},d_A,d_B)$ TS code such that
	\begin{equation}\label{GV-time-sharing}
	A^{TS}=G(n_A,d_A):=A^{TS}_G,\  B^{TS}=G(n_B,d_B):=B^{TS}_G,
	\end{equation}
	where $G(n,d)$ is defined in \eqref{G}.
\end{lemma}

Due to the absence of a concise formula for the parameters of the
optimal TS codes, in the following we employ the size provided by the GV bounds, $A^{TS}_G$ and  $B^{TS}_G$, for theoretical comparison with UEP codes. Note that although codes achieving
GV bound are not necessarily optimal, they exhibit asymptotic optimality and can be comparable with the optimal codes \cite{GV}, \cite{GV1}.
\subsection{Theoretical comparison of UEP and TS}\label{3.2}
Throughout this section, we always assume $n=n_A+n_B$ and compare the $(n, A, B,d_A,d_B)$-UEP code with $(n_A,n_B, A^{TS}, B, d_A,d_B)$-TS code, \textit{i.e.} we care about the increment in size $|\mathcal{A}|$ while keeping all other parameters coincident. 

According to the improved GV bound for UEP stated in Theorem \ref{Best theorem}, we claim that there always exists a UEP codes with larger size $|\mathcal{A}|$ than the $TS$ codes with size given by the GV bounds $A_G^{TS}$ and $B_G^{TS}$ in \eqref{GV-time-sharing}. In fact, the ``cubes" $\{D_i:=C^A_{TS}(m^{\mathcal{A}}_i)\times F_2^{n_B}: i=1,2,\cdots,A_G^{TS}\}\in\mathcal{D}(n, B, d_A,d_B)$. Therefore, we know that there exists an $(n,A,B_G^{TS},d_A,d_B)$-UEP code with 
\begin{equation}
A \geq A_G^{TS}.
\end{equation} 
The  improvement comes from that Theorem \ref{Best theorem}
constructs part of the codewords in a centralized way, and then fills in the remaining gaps as much as possible, while TS does not execute the second step.

In the following, we show that under certain conditions, the asymptotic rate gain of UEP over TS is non-vanishing. For TS code, we  take the optimal length allocation $n_A, n_B$ according to GV bound
\eqref{GV-time-sharing} when $B_G^{TS}=B$, \textit{i.e.},
$
n_B:=\min\{m: \frac{2^m}{V(m,d_B-1)}>B-1\},
$
$n_A=n-n_B,$ and $ A^{TS} = A_G^{TS}$.
We denote
$
\alpha^*:=\frac{n_A}{n}
$
be the optimal code length allocation.
 Then the achievable size $A$ in Corollary \ref{GV-2-1} satisfies
\begin{equation}
A\approx  \frac{2^{n_A}V(n_B, d_B-1)- V(n,d_B-1)}{V(n,d_A-1)}\approx \frac{2^{n_A}V(n_B, d_B-1)}{V(n,d_A-1)}.
\end{equation}
Thus
\begin{equation}\label{fix d}
\begin{aligned}
\frac{A}{A^{TS}}&\approx \frac{V(n_A,d_A-1)V(n_B, d_B-1)}{V(n,d_A-1)}\\
&\overset{(a)}{\approx} \frac{(\alpha^*)^{d_A-1}(1-\alpha^*)^{d_B-1}}{(d_B-1)!}n^{d_B-1},
\end{aligned}
\end{equation}
where $(a)$ follows from $V(n,r+1)=\frac{n^r}{r!}(1+O(1/n))$, $\alpha^*=\alpha^*(n,d_B,B):=n_A/n$. Obviously,  if
\begin{equation}
d_B<d_A<\log_\alpha^* \left( \frac{(d_B-1)!}{[n(1-\alpha^*)]^{d_B-1}}\right)+1 ,
\end{equation}
then we conclude that there is a non-vanishing rate gain of UEP over TS.  If we assume $d_A=\beta_A n, d_B=\beta_B n$ where $1/2\alpha^*>\beta_A>\beta_B$, we can deduce the following size comparison.
\begin{theorem}\label{size compression of BSC under new bound of ball}
	Given $B$. Then for any $(\beta_A,\beta_B)$ satisfying
	$
	\beta_A\le \frac{1}{2}\alpha^*,\;\; \beta_B\le \min\{\frac{1}{2}(1-\alpha^*), \beta_A\},
	$
	and
	\begin{equation}\label{condition of gain}
	\alpha^* h(\frac{\beta_A}{\alpha^*})+(1-\alpha^*)h(\frac{\beta_B}{1-\alpha^*})-h(\beta_A)>0,
	\end{equation}
	there exists $n_0=n_0(\alpha^*, \beta_A,\beta_B)$ \textit{s.t.} $\forall n>n_0$, there exists a $(n, A, B, \beta_A n, \beta_B n)$-UEP code and a constant $a>1$ independent of $n$ such that $A>a^nA^{TS}$, where $A^{TS}$ is the parameter of the TS code $(\alpha^* n, (1-\alpha^*)n, A^{TS}, B, \beta_A n, \beta_B n)$, which achieves GV bound.
\end{theorem}
\pf It is well-known that (\textit{e.g.}, see \cite{yehezkeally2022bounds})
		\begin{equation}
		\frac{1}{n+1}2^{n h(r/n)}\le V(n,r)\le 2^{n h(r/n)}, \forall r\leq \frac{n}{2}.
		\end{equation}
It follows that
	\begin{equation}
	\begin{aligned}
	\frac{A}{A^{TS}}&\approx \frac{V(n_A,d_A-1)V(n_B,d_B-1)}{V(n,d_A-1)}\\
	&\ge  \frac{1}{\alpha^* (1-\alpha^*)n^2}2^{n[\alpha^* h(\frac{\beta_A}{\alpha^*})+(1-\alpha^*)h(\frac{\beta_B}{1-\alpha^*})-h(\beta_A)]},
	\end{aligned}
	\end{equation}
which completes the proof.
	\e
\begin{remark}
Using the monotonicity the function $f(\alpha^*,\beta_A,\beta_B):=h(\frac{\beta_A}{\alpha^*})+(1-\alpha^*)h(\frac{\beta_B}{1-\alpha^*})-h(\beta_A)$, we deduce that if $\alpha^*(n,d_B,B)\ge0.1$, then \eqref{condition of gain} holds for all $(\beta_A,\beta_B)$ satisfying
	\begin{equation}
	0<\frac{\beta_A}{2}\le \beta_B<\beta_A\le \frac{\alpha^*}{3}.
	\end{equation}
\end{remark}
\begin{remark}
Note that $h(\cdot)$ is increasing, which implies the conditions in Theorem \ref{size compression of BSC under new bound of ball} holds if	
	\begin{equation}
	0<(1-\alpha^*)\beta_A\le \beta_B\le \beta_A\le \frac{1}{2}\alpha^*.
	\end{equation}

\end{remark}

\section{SIMULATION}\label{section5}
Table \ref{Simulation for  Bounds} presents the achievable length of UEP and TS codes with short code lengths. Specifically, $n^{TS}_G$ is an achievable code length obtained by GV bound for TS as in \eqref{GV-time-sharing},  while $n^{TS}_b$ is the minimum code length of TS code according to codetables.de \cite{Grassl:codetables}. $n^U$ is deduced according to Theorem \ref{Best theorem} where $D_i$s are chosen as  the ``cubes" $\{D_i:=C^A_{TS}(m^{\mathcal{A}}_i)\times F_2^{n_B}\}$ with $n_B$ satisfying
$$
B\le \frac{2^{n_B}}{V(n_B,d_B-1)}.
$$
$n^u$ is the length of the optimal LUEP given in \cite{twotopic}.
	\begin{table}[t]
	\centering
	
	\caption{Simulation for  Bounds}
	\begin{tabular}{|c|c|c|c|c|c|c|c|}
		\hline
		$\log_2 A$&$\log_2 B$&$d_A$&$d_B$&$n^{TS}_G$  &$n^{TS}_b$&$n^U$& $n^u$\\
				\hline
	2&3&5&4&22&15&16&11\\
	\hline
	2&4&5&4&24&16&18&12\\
	\hline
	2&3&6&4&24&16&17&12\\
	\hline
	4&5&3&2&19&13&16&12\\
	\hline
	4&6&4&2&23&15&14&14\\
	\hline
	2&4&7&4&28&19&20&15\\
	\hline
	4&7&4&2&24&16&15&15\\
	\hline
	4&8&3&2&22&16&20&15\\
	\hline
	\end{tabular}
	\label{Simulation for  Bounds}
\end{table}

It is evident that $n^u$ surpasses $n^{TS}_b$ for all cases, which indicates the superiority of UEP over TS for short codes. Furthermore, $n^U$ is close to $n^u$ when $d_B=2$, which implies that this achievability bound is tight when $d_B$ is small.

For longer code length, we summarize our achievability bounds as follows. By choosing $D_i$s as ``cubes" and using Theorem \ref{Best theorem}, we obtain an achievability size 
\begin{equation}
A_1=A^{TS}_{G}+\frac{2^n - (B-1)V(n,d_B-1)-A^{TS}_{G}I_1}{BV(n,d_A-1)}\vee 0.
\end{equation}
where $I_1=I(n,B,d_B,d_A-1)$.
If we choose $D_i$s as balls, then Theorem \ref{Best theorem} and \eqref{bound of MS} imply an  achievability size
\begin{equation}
A_2=M_S^L+\frac{2^n - (B-1)V(n,d_B-1)-M_S^LI_1}{BV(n,d_A-1)}\vee 0,
\end{equation}
where $M_S^L=2^n/V(n,r_S)$ is a lower bound of $M_S$.
Furthermore, if we use the enlargement bound and Theorem \ref{GV-ball}, another  achievability size is obtained as
\begin{equation}
A_3=M_S^L+\frac{2^n - (B-1)V(n,d_B-1)-I_2}{BV(n,d_A-1)}\vee 0,
\end{equation}
where $I_2=I(n,M_S^L,2r_S+1,r_V+d_A-1)$.

\begin{figure}[t]
	\centering
	{\subfigure
		{\includegraphics[width=0.24\textwidth]{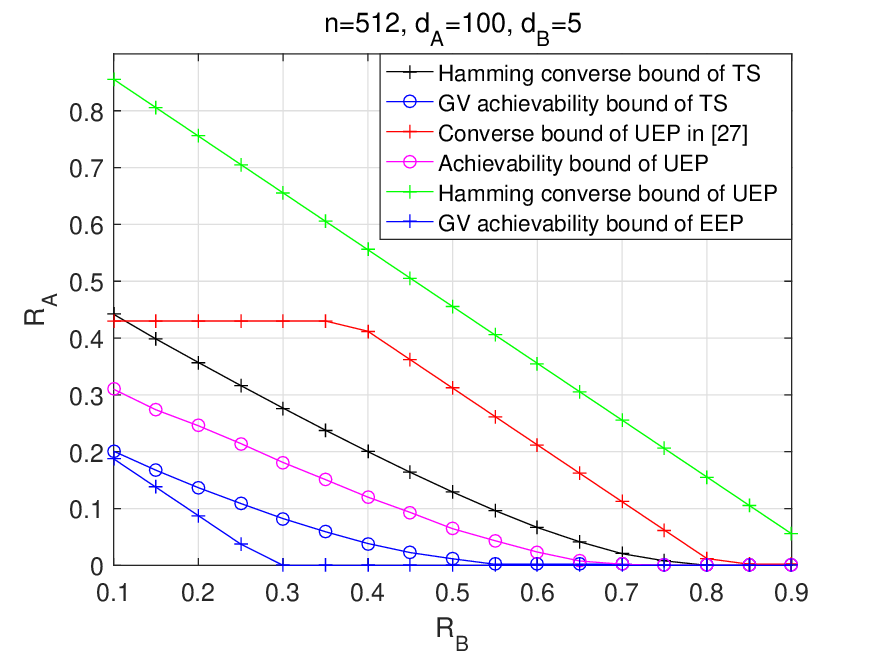}}
		\subfigure
		{\includegraphics[width=0.24\textwidth]{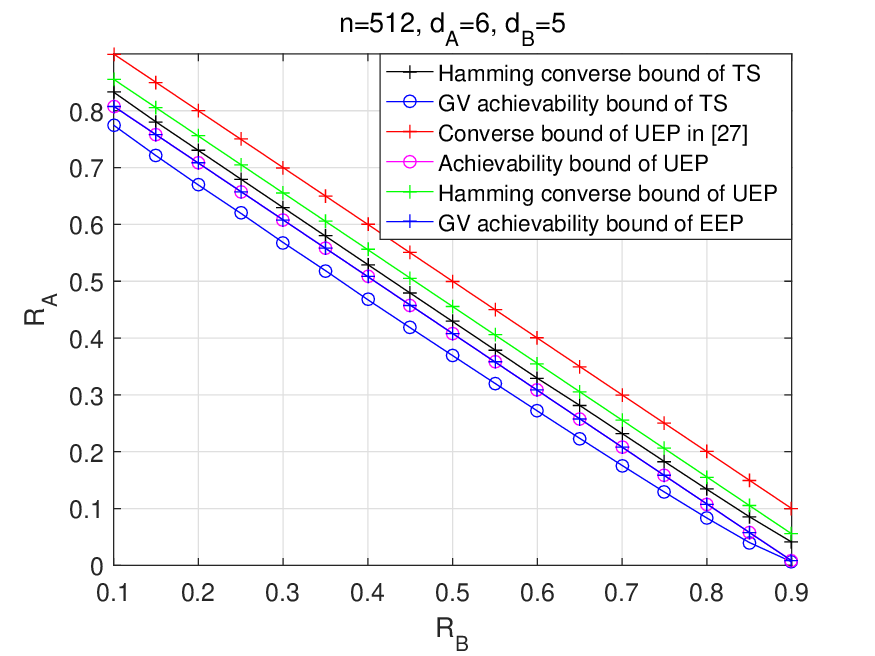}}}
	\caption{Comparison on  Several Bounds of TS and UEP Codes}
	\label{Simulation of Several Bounds}
\end{figure}

The  rate obtained by   bounds  with longer code lengths are shown in Fig. \ref{Simulation of Several Bounds}, where the achievability bound of UEP is the maximum of our results,
\textit{i.e.} $\frac{\log_2 \max\{A_1, A_2, A_3    \}}{n}.$
The converse bound of UEP is from \cite{upper_of_vol}. The Hamming (converse) bound of UEP is
\begin{equation}
A\le \frac{2^n}{BV(n,\lceil (d_B-1)/2\rceil)},
\end{equation}
by trivially dropping the code distance  requirement of $\mathcal{A}$  from $d_A$ to $d_B$. 

The achievability bound 
\begin{equation}
A=\frac{2^n}{BV(n_A,d_A-1)}
\end{equation}
of Equal-Error Protection (EEP) Code (defined as the UEP code with $d_1 = d_2 =\cdots = d_m$)  \cite{Ef}  is also included  
in the comparison. The EEP elevates the code spacing requirement to be consistent, which corresponds to the UEP code  in Definition \ref{de of UEP} with $d_1=d_2=\cdots=d_m$. Although it will lose the rate due to the increased code distance, compared with TS, it constructs the codeword in the whole space without spatial segmentation, thus the advantages and disadvantages of the two strategies are not obvious (see Fig. \ref{Simulation of Several Bounds}).

From the simulations, we find that  when $d_B\ll d_A$, UEP can obtain significant advantages. 	This is because UEP gathers  $B$ codewords with distance requirement $d_B$ together, and thus the intersection of the balls with  radius $d_A$ is  large. Therefore, there is more space left to place more codewords and the gain is obtained comparing with TS and EEP.	It is notable that the achievability bounds are not  coincide for $R_B=0$, since in TS we always ask $n_B\ge d_B$. 
When $d_B$ is close to $d_A$, the  TS awkwardly combines two codes with similar protection levels independently and thus loses the code length gain. Furthermore, UEP does not show significant improvement compared to EEP when $d_B \approx d_A$, indicating that UEP is not necessary in this case.

\section{Conclusion}\label{section6}


In this paper, we obtain the achievability bounds of  UEP  codes, and prove that UEP is superior to TS under certain conditions. We first consider the number of  disjoint connected sets in the space $\mathbb{F}_2^n$ and utilize a tighter estimation to the volume of union balls, such as intersection bound and enlargement bound, and then fill in the remaining gaps as much as possible. It is demonstrated that the UEP codes obtained by the above procedures achieve significant gains comparing with TS and EEP when $d_B\ll d_A$.

\begin{appendix}
\subsection{Proof of Remark \ref{LUEP}}\label{pf of LUEP}
We denote the information as $u\in (u_1^{k_A }, u_{k_A +1}^{k_A +k_B })\in F_2^{k_A +k_B }$, where $u_1^{k_A }$ represent the messages in $\mathcal{A}$ and $u_{k_A +1}^{k_A +k_B }$ represent the messages in $\mathcal{B}$.

Now we consider random matrix $G$ with \textit{i.i.d.} Bernoulli($\frac{1}{2}$) entries. Then fix any $u\in F_2^{k_A +k_B }\setminus\{0\}$, vector $uG$ also has \textit{i.i.d.} Bernoulli($\frac{1}{2}$) entries. Thus the event $A_{u,d}:=\{\rho_H(uG)\le d\}$ has probability
\begin{equation}
\P(A_{u,d})=\frac{V(n,d)}{2^n}, \forall u\neq 0.
\end{equation}

Since we need $d_A$ minimum distance for $u_1^{k_A}$ and $d_B$ minimum distance for $u_{k_A +1}^{k_A +k_B }$, the feasible event should be $\Theta:=\left( (\bigcup_{u\in S_1} A_{u,d_A})\bigcup(\bigcup_{u\in S_2} A_{u,d_B})\right)^c$, where
\begin{equation}
S_1:=\{u: u_1^{k_A}\neq 0\},\;\; S_2:=\{u: u_{k_A +1}^{k_A +k_B }\neq 0\}.
\end{equation}
Note that 
\begin{align}
\P(\Theta)&=1-\P\left( (\bigcup_{u\in S_1} A_{u,d_A})\bigcup(\bigcup_{u\in S_2} A_{u,d_B})\right)\\
&\overset{(a)}{=} 1-\P\left( (\bigcup_{u\in S_1} A_{u,d_A})\bigcup(\bigcup_{u\in S_1^c\cap S_2} A_{u,d_B})\right)\\
&\ge 1-\left( \sum_{u\in S_1}\P(A_{u,d_A})+\sum_{u\in S_1^c\cap S_2}\P(u,d_B)\right)\\
&=1-\frac{\mu(S_1)V(n,d_A)+\mu(S_1^c\cap S_2)V(n,d_B)}{2^n}\\
&=1-\frac{B(A-1)V(n,d_A)+(B-1)V(n,d_B)}{2^n}\\
&\overset{(b)}{>}0,\label{>0}
\end{align}
where $(a)$ comes from $d_A>d_B$ and $(b)$ comes from \eqref{union size}. Clearly, \eqref{>0} means the existence of generate matrix $G_0$ \textit{s.t.} 
\begin{equation}
\rho_H(uG_0)\ge d_A,\; \forall u\in S_1;\;\; \rho_H(uG_0)\ge d_B,\; \forall u\in S_2,
\end{equation}
\textit{i.e.} there exists a LUEP code that satisfies the requirements of code distance  and code size.
\e

\subsection{Proof of $g(x,y)>0,\forall x,y\in(0,1/2)$}\label{g>0}
\pf Fix any $x_0\in(0,1/2)$, then $g_0(y):=g(x_0,y)$ satsifying that $g_0'(y)=h'(y)-h'(x+y)>0$ since $h''(x)<0,\forall x\in(0,1)$, then we conclude $g_0(y)>0,\forall y\in(0,1/2)$ by $g_0(0)=0$. 
\e

\end{appendix}

\end{document}